\documentclass[aps,pra,preprint,eqsecnum]{revtex4}

\usepackage{graphicx}
\usepackage{epsfig}

\newcommand{\be}{\begin{equation}}
\newcommand{\ee}{\end{equation}}
\newcommand{\bea}{\begin{eqnarray}}
\newcommand{\eea}{\end{eqnarray}}

\newcommand{\ra}{\rightarrow}
\newcommand{\ua}{\uparrow}
\newcommand{\da}{\downarrow}
\newcommand{\lra}{\leftrightarrow}

\newcommand{\RR}{\rangle}
\newcommand{\LL}{\langle}

\newcommand{\RL}{\RR\LL}

\begin{document}

\title{Distant entanglement with nearest neighbor interactions}

\author{Gavin K. Brennen}
\affiliation{National Institute of Standards and Technology, Gaithersburg, MD  20899-8423, USA}

\date{\today}

\begin{abstract}

Preparing many body entangled states efficiently using available interactions is a challenging task.  One solution may be to couple a system collectively with a probe that leaves residual entanglement in the system.  We investigate the entanglement produced between two possibly distant
qubits 1 and 2 that interact locally with a third qubit 3 under unitary evolution
generated by pairwise Hamiltonians.  For the case where
the Hamiltonians commute, relevant to certain quantum nondemolition measurements, the entanglement between qubits 1 and 2 is calculated
explicitly for several classes of initial states and compared with the case of
noncommuting interaction Hamiltonians.  This analysis can be helpful to identify preferable
physical system interactions for entangled state synthesis.
\end{abstract}

\pacs{}
\maketitle

\section{Introduction}

Recent advances in the preparation and manipulation of many body entangled states has
promoted keen interest in characterizing entanglement as a resource for quantum
information processing with applications to such areas as precision measurement, quantum communications, and computation \cite{nielsenbook}.  Given a many body system Hamiltonian it is possible to construct designer entangled states using only single body and at least one two body interaction, however, it is difficult in general to find the optimal sequence of control operations to prepare designer entangled states.  Often the number of control operations needed scales poorly with the number of bodies, or interaction with the environment introduces significant decoherence during the preparation.  One candidate for efficient production of multiparticle entanglement is to collectively interact members of a system with a probe that can be measured after the interaction.  This has been proposed in the context of quantum nondemolition (QND) measurements wherein a probe $p$ interacts with the many particle system $s$ via an interaction Hamiltonian $H_{ps}$.  If an observable $M_{s}$ of the system commutes with $H_{ps}$ and is a constant of motion in the absence of the interaction, then measurement of an observable $M_{p}$ that does not commute with with $H_{ps}$ constitutes a QND measurement of $M_{s}$.  Such a technique can be used to prepare many body entangled states of the system.

As an example, the interaction of a single frequency mode (two polarization mode) laser field in a coherent state with a collection of two level atoms can be modeled in a truncated Fock space as a $2n$-dimensional system interacting with a collection of qubits.
A nondemolition measurement of the $z$ component of collective spin of the atoms, $S_{z}=\sum_{i} \sigma^{i}_{z}$,
is made by interacting the field with the atoms collectively via a pairwise commuting interaction $H=g\sum_{i} J_{z}\sigma^{i}_{z}$, where  $J_{z}=1/2(a^{\dagger}_{+}a_{+}-a^{\dagger}_{-}a_{-})$ is the angular momentum component along $z$ in the Schwinger representation where $a_{\pm}$ are the annihilation operators for each polarization mode.  A subsequent measurement of a canonically conjugate variable of the field such as  $J_{x}=1/2(a^{\dagger}_{+}a_{-}+a^{\dagger}_{-}a_{+})$ completes the QND measurement.  Such an interaction can be realized as the polarization
dependent AC
stark shift on the atoms \cite{takahashi}.  Under unitary evolution the atoms become entangled with the field and by measuring the field it has experimentally been shown that one can prepare entangled atomic spin squeezed states
\cite{kuzmich}.
To see this consider 2 two level atoms, 1 and 2, that interact with a photonic qubit 3 storing information in the polarization degrees of freedom.  Let 3 be prepared in the state $|J_{x}=J=\frac{1}{2}\RR=1/\sqrt{2}(a^{\dagger}_{+}+a^{\dagger}_{-})|0\RR$ where $|0\RR$ is the vacuum state, and the two qubits be prepared in a product state with both spins oriented along \emph{x}, or written in the coupled basis $|S,S_{z}\RR$,
\begin{equation}
\begin{array}{lll}
|S_{x}=1\RR&=&|\ua_{x}\RR|\ua_{x}\RR\\
&=&\frac{1}{2}(|1,-1\RR+\sqrt{2}|1,0\RR+|1,1\RR.\;
\end{array}
\end{equation}
Under evolution by the Hamiltonian $H=g J_{z} \otimes S_{z}$ the state evolves to,
\begin{equation}
\begin{array}{lll}
U(t)|S_{x}=1\RR|J_{x}=1\RR&=&\frac{1}{2}(\cos(\frac{gt}{2})(|1,-1\RR+|1,1\RR)|J_{x}=\frac{1}{2}\RR+\sqrt{2}|1,0\RR|J_{x}=\frac{1}{2}\RR\\
& &+i\sin(\frac{gt}{2})(|1,-1\RR-|1,1\RR)|J_{x}=-\frac{1}{2}\RR).\;
\end{array}
\end{equation}   
After interaction times satisfying $gt=\pi(2m+1)$, for $m$ any natural number, a measurement of 3 along $|J_{x}=1\RR$ yields the conditional maximally entangled state for 1 and 2: $|1,0\RR=|\psi^{+}\RR=1/\sqrt{2}(|\ua_{z}\RR|\da_{z}\RR+|\ua_{z}\RR|\da_{z}\RR)$.

It would be useful to know what kind of entanglement is produced within the collective sample before a measurement is made on the probe.  Such knowledge could indicate what type of generalized measurement is optimal to prepare the desired state of the system.  It is generally quite difficult to characterize the dynamics of entanglement between many bodies.  Indeed, even a tractable quantification of entanglement between one qubit and an $n$ dimensional system for $n\geq2$ has yet to be found in the general \cite{horodecki}.  We study the simplest model possible namely the dynamics of entanglement between two qubits 1 and 2 that interact pairwise with a third qubit 3.  We show that there are some easily identified constraints to the dynamics of entanglement starting from various initial pure states of the three body system in the case that the interaction Hamiltonians $H_{13}$ and $H_{23}$ commute.  This is the regime encountered in QND measurements of the type outlined above.  In the case of noncommuting Hamiltonians the dynamics are less restrictive but can be characterized by certain symmetries of the interaction.

\section{Commuting interactions}

We begin with a system of two qubits $1$ and $2$ interacting with a third qubit $3$ via
pairwise interaction Hamiltonians:
\begin{equation}
\begin{array}{lll}
H_{13}&=& \displaystyle{\sum_{i,j=1}^{3}} \alpha_{ij} \sigma^{1}_{i} \otimes \sigma^{3}_{j}
+ \sum_{k=1}^{3} \alpha^{\prime}_{k} \sigma^{1}_{k} \otimes \mbox{\boldmath $1$}^{3}
+ \sum_{k=1}^{3} \alpha^{\prime\prime}_{k} \mbox{\boldmath $1$}^{1} \otimes \sigma^{3}_{k},\\
H_{23}&=& \displaystyle{\sum_{i,j=1}^{3}} \beta_{ij} \sigma^{2}_{i} \otimes \sigma^{3}_{j}
+ \sum_{k=1}^{3} \beta^{\prime}_{k} \sigma^{2}_{k} \otimes \mbox{\boldmath $1$}^{3}
+ \sum_{k=1}^{3} \beta^{\prime\prime}_{k} \mbox{\boldmath $1$}^{2} \otimes \sigma^{3}_{k},\;
\end{array}
\end{equation}
where $\sigma^{m}_{n}$ denotes the Pauli operator
$\sigma_{(1,2,3)}\equiv \sigma_{(x,y,z)},$ acting on the space of qubit $m$.
If $[H_{13},H_{23}]=0$, then it is straightforward to show that the Hamiltonians must
assume the following form:
\begin{equation}
\begin{array}{lll}
H_{13}&=& | \vec{\alpha}| \sigma^{1}_{\vec{\alpha}} \otimes
\sigma^{3}_{j}+| \vec{ \alpha^{\prime}}| \sigma^{1}_{ \vec{ \alpha^{\prime}}}
\otimes \mbox{\boldmath $1$}^{3}+\alpha^{\prime\prime} \mbox{\boldmath $1$}^{1}
\otimes \sigma^{3}_{j}\\
H_{23}&=& | \vec{\beta}| \sigma^{2}_{\vec{\beta}} \otimes
\sigma^{3}_{j}+| \vec{ \beta^{\prime}}| \sigma^{2}_{ \vec{ \beta^{\prime}}}
\otimes \mbox{\boldmath $1$}^{3}+\beta^{\prime\prime} \mbox{\boldmath $1$}^{2}
\otimes \sigma^{3}_{j}, \;
\end{array}
\end{equation}
where $\sigma^{m}_{\vec{\gamma}} \equiv 1/|\vec{\gamma}| (\gamma_{1} \sigma^{m}_{1}+
\gamma_{2} \sigma^{m}_{2}+\gamma_{3} \sigma^{m}_{3})$.
The second two terms in each of $H_{13}$ and $H_{23}$ commute with all other terms
and have the effect of local operations acting after the evolution generated by the entangling
Hamiltonians:  $H_{13}^{\prime}=
| \vec{\alpha}| \sigma^{1}_{\vec{\alpha}} \otimes\sigma^{3}_{j},
H_{23}^{\prime}=| \vec{\beta}| \sigma^{2}_{\vec{\beta}} \otimes
\sigma^{3}_{j}$.  Local unitary operations leave the entanglement between
qubits 1 and 2 invariant and as such we consider only the reduced state
of qubits 1 and 2 after joint evolution by the unitary
$U(t)=e^{-i(H_{13}^{\prime}+H_{23}^{\prime})t}$.

An arbitrary pure state of qubits 1,2 and 3 can be written in
the eigenbasis of $\sigma^{1}_{\vec{\alpha}}\otimes\sigma^{2}_{\vec{\beta}}
\otimes\sigma^{3}_{j}$ which is the simultaneous eigenbasis of $H_{13},H_{23}$, as
\begin{equation}
\ | \Psi\RR= \sum a_{\pm\pm\pm} | \pm\RR_{\vec{\alpha}} \otimes\
| \pm\RR_{\vec{\beta}} \otimes | \pm\RR_{j} \equiv \sum a_{\pm\pm\pm}
| \displaystyle{\pm\pm}\pm\RR. \;
\end{equation}
After evolution for a time $t$, the state is compactly written
\begin{equation}
U(t)| \Psi\RR=\displaystyle{\sum_{m_{1},m_{2},m_{3}\in\{-1,+1\}}} a_{m_{1},m_{2},m_{3}}
e^{-i(m_{1}m_{3}| \vec{\alpha}|+m_{2}m_{3}| \vec{\beta}|)t}|m_{1}m_{2}m_{3}\RR, \;
\label{genstate}
\end{equation}
and the reduced state of qubits 1 and 2 is
\begin{equation}
\begin{array}{lll}
\rho_{12}(t)&=&Tr_{3}[U(t)| \Psi\RL\Psi|U^{\dagger}(t)]\\
&=&\displaystyle{\sum_{all \ m \in\{-1,+1\}}}
a_{m_{1},m_{2},m_{3}}a_{m_{1}^{\prime},m_{2}^{\prime},m_{3}}^{*}
e^{-i((m_{1}-m_{1}^{\prime})m_{3}| \vec{\alpha}|+(m_{2}-m_{2}^{\prime})m_{3}| \vec{\beta}|)t}
|m_{1}m_{2}\RL m_{1}^{\prime}m_{2}^{\prime}|.\;
\end{array}
\end{equation}

Wootters \cite{wootters} has given an explicit formula to calculate the entanglement of formation
$\mathcal{E}(\rho)$ for an arbitrary state of two qubits in terms of the tangle defined as
$\tau_{12}=[max\{\lambda_{1}-\lambda_{2}-\lambda_{3}-\lambda_{4}\}]^{2}$
where the $\lambda_{i}$ are the square roots of the eigenvalues, in decreasing order, of
$\rho_{12}\tilde{\rho}_{12}$, where $\tilde{\rho}_{12}$ is the spin-flipped version of $\rho_{12}$,
$\tilde{\rho}_{12}=\sigma_{y} \otimes \sigma_{y} \rho^{*}_{12} \sigma_{y} \otimes \sigma_{y}$.
The entanglement of formation is a monotonically increasing function of the tangle given by
$\mathcal{E}$$(\rho)=h(1/2 + 1/2 \sqrt{1-\tau})$ where $h(x)$ is the binary
entropy function.  A normalized pure state of
three qubits will generally have $15$ independent real parameters ($5$ of which are invariant
under local unitaries \cite{linden}) making it cumbersome to calculate this quantity.  We
focus on simply determining how the entanglement between 1 and 2 changes after joint
evolution for different classes of initial states.

\subsection{Fully separable initial states}
Consider the case when the initial state is pure and fully separable,
\begin{equation}
| \Psi\RR=R_{1}\otimes R_{2}\otimes R_{3}| \displaystyle{++}+\RR, \;
\end{equation}
where the local unitary operators are written $R_{k}=e^{-i \gamma_{k} \sigma^{k}_{\vec{\gamma}_{k}}}$.
Quite generally,
\begin{equation}
\begin{array}{lll}
R_{3}|+\RR&=&c|+\RR+d|-\RR,\\
R_{3}|-\RR&=&d^{*}|+\RR-c^{*}|-\RR, \;
\end{array}
\label{rot}
\end{equation}
where $|c|^{2}+|d|^{2}=1$, and we consider the action of $U(t)$ on the states
$R_{1}\otimes R_{2}|\displaystyle{++}\raisebox{-.2ex}{\mbox{$\pm$}}\RR$ separately.
Because $[H_{13},H_{23}]=0$, we write
$U(t)=e^{-iH_{13}^{\prime}t}e^{-iH_{23}^{\prime}t}=U_{13}(t)U_{23}(t)$.  Using the fact that
the operators in the generating Hamiltonians are unitary and hermitian the
unitaries can be expanded as
\begin{equation}
\begin{array}{lll}
U_{13}(t)&=&\cos(|\vec{\alpha}|t) \mbox{\boldmath$1$}-i\sin(|\vec{\alpha}|t) \sigma^{1}_{\vec{\alpha}}
\otimes\sigma^{3}_{j}\\
U_{23}(t)&=&\cos(|\vec{\beta}|t) \mbox{\boldmath$1$}-i\sin(|\vec{\beta}|t) \sigma^{2}_{\vec{\beta}}
\otimes\sigma^{3}_{j},\;
\end{array}
\end{equation}
and similarly, $R_{k}=\cos(\gamma_{k})-i\sin(\gamma_{k})\sigma^{k}_{\vec{\gamma}_{k}}$.
The evolution can be divided into two pieces,
\begin{equation}
\begin{array}{lll}
U_{13}(t)R_{1}|\displaystyle{++}\raisebox{-.2ex}{\mbox{$\pm$}}\RR&=&(\cos(|\vec{\alpha}|t)\cos(\gamma_{1})
\mbox{\boldmath$1$}
-i\cos(|\vec{\alpha}|t)\sin(\gamma_{1})\sigma^{1}_{\vec{\gamma_{1}}} \\
& &-i\sin(|\vec{\alpha}|t)\cos(\gamma_{1})\sigma^{1}_{\vec{\gamma_{1}}}
\otimes\sigma^{3}_{j}-\sin(|\vec{\alpha}|t)\sin(\gamma_{1})\sigma^{1}_{\vec{\alpha}}
\sigma^{1}_{\vec{\gamma_{1}}}\otimes\sigma^{3}_{j})| \displaystyle{++}\raisebox{-.2ex}{\mbox{$\pm$}}\RR \\
&=&e^{\mp i|\vec{\alpha}|t\sigma^{1}_{\vec{\alpha}}}R_{1}|
\displaystyle{++}\raisebox{-.2ex}{\mbox{$\pm$}}\RR \\
&\equiv&V_{1_{\pm}}(t)| \displaystyle{++}\raisebox{-.2ex}{\mbox{$\pm$}}\RR,\\
\\
U_{23}(t)R_{2}|\displaystyle{++}\raisebox{-.2ex}{\mbox{$\pm$}}\RR&=&
e^{\mp i|\vec{\beta}|t\sigma^{2}_{\vec{\beta}}}R_{2}| \displaystyle{++}\raisebox{-.2ex}{\mbox{$\pm$}}\RR \\
&\equiv&V_{2_{\pm}}(t)|\displaystyle{++}\raisebox{-.2ex}{\mbox{$\pm$}}\RR. \;
\end{array}
\label{singleops}
\end{equation}
The evolved state of the three qubit system is generally entangled,
\begin{equation}
U(t)| \Psi\RR=c V_{1_{+}}(t)\otimes V_{2_{+}}(t)|\displaystyle{++}+\RR
+d V_{1_{-}}(t)\otimes V_{2_{-}}(t)|\displaystyle{++}-\RR. \;
\end{equation}
Tracing over the third qubit, the reduced state of qubits 1 and 2 is
\begin{equation}
\begin{array}{lll}
\rho_{12}(t)&=&|c|^{2} V_{1_{+}}(t)\otimes V_{2_{+}}(t)| \displaystyle{++}\RL\displaystyle{++}
|V_{1_{+}}(t)^{\dagger}\otimes V_{2_{+}}(t)^{\dagger}\\
& &+|d|^{2} V_{1_{-}}(t)\otimes V_{2_{-}}(t)|\displaystyle{++}\RL\displaystyle{++}|
V_{1_{-}}(t)^{\dagger}\otimes V_{2_{-}}(t)^{\dagger}; \;
\end{array}
\end{equation}
a convex sum over separable states and therefore seperable,
$\mathcal{E}$$(\rho_{12}(t))=$$\mathcal{E}$$(\rho_{12}(0))=0$.

The mapping on the reduced state $\rho_{12}=Tr_{3}(| \Psi\RL\Psi |)$ can also be written
\begin{equation}
\rho_{12}(t)=\sum V_{1_{\pm}}(t)R_{1}^{\dagger} \otimes V_{2_{\pm}}(t)R_{2}^{\dagger} {_{3}
\LL}\pm| \Psi \RL \Psi|\pm\RR_{3}
V_{1_{\pm}}(t)^{\dagger}R_{1} \otimes V_{2_{\pm}}(t)^{\dagger}R_{2}, \;
\end{equation}
where $R_{1}$ and $R_{2}$ transform the basis states given in the expansion of $| \Psi\RR$
to the eigenbasis of the interaction.
In the case that the state of qubits 1 and 2 is initially uncorrelated with qubit
3, i.e. $| \Psi\RR=| \chi\RR_{12} \otimes | \phi\RR_{3}$, then it is convenient to express
the evolution of the reduced state $\rho_{12}(t)$
in Krauss form \cite{krauss}
\begin{equation}
\rho_{12}(t)=\sum A_{\pm}(t)| \chi\RL \chi|A_{\pm}(t)^{\dagger}, \;
\end{equation}
where the Krauss operators are defined
\begin{equation}
A_{\pm}(t)= {_{3}\LL}\pm|U(t)| \Psi\RR_{3}
={_{3}\LL}\pm| \phi\RR_{3} V_{1_{\pm}}(t)R_{1}^{\dagger} \otimes V_{2_{\pm}}(t)R_{2}^{\dagger}, \;
\end{equation}
satisfying the trace perserving condition:
$\sum A_{\pm}(t)^{\dagger}A_{\pm}(t)=\mathbf{1}$.
One could also find a Krauss operator expansion for the case where the initial
state of the joint system of qubits 1,2 and qubit 3 are entangled.  This would require finding the
unitary that maps an initial unentangled state $| \chi\RR_{12} \otimes | \phi\RR_{3}$ to the entangled
one.

\subsection{Entangled initial states}
Here we consider the action of the unitary evolution on initially entangled
states from various classes.

\subsubsection{Bipartite entanglement between 1 and 2}
Let the initial state be
\begin{equation}
\begin{array}{lll}
| \Psi\RR&=&| \chi\RR_{12}| \phi\RR_{3}\\
&=&(a|0\RR_{1}|0\RR_{2}+b|1\RR_{1}|1\RR_{2})| \phi\RR_{3},\
\end{array}
\end{equation}
where we have expanding the entangled state $| \chi\RR$ between qubits 1 and 2
in a Schmidt basis with real parameters $a,b$.
This state has an initial tangle between qubits 1 and 2 $\tau_{12}(|\chi\RR)=4(ab)^{2})$.
Using $| \phi\RR=R_{3}|+\RR$ with $R_{3}$ given in Eq. \ref{rot},
the state $| \Psi\RR$ can be written as a linear combination of eigenstates of
$\sigma^{1}_{\vec{\alpha}}\otimes\sigma^{2}_{\vec{\beta}}\otimes\sigma^{3}_{j}$:
\begin{equation}
| \Psi\RR=c R_{1} \otimes R_{2}(a|\displaystyle{++}\RR+b|\displaystyle{--}\RR)_{12}|+\RR_{3}
+d R_{1} \otimes R_{2}(a|\displaystyle{++}\RR+b|\displaystyle{--}\RR)_{12}|-\RR_{3}.\;
\end{equation}

The Krauss operators are therefore,
\begin{equation}
\begin{array}{llll}
A_{+}(t)&=&c V_{1_{+}}(t)R_{1}^{\dagger} \otimes V_{2_{+}}(t)R_{2}^{\dagger} \\
A_{-}(t)&=&d V_{1_{-}}(t)R_{1}^{\dagger} \otimes V_{2_{-}}(t)R_{2}^{\dagger},\;
\end{array}
\end{equation}
with $V_{j_{\pm}}$ given by Eq. \ref{singleops} with the appropriate rotation operators
$R_{k}$ for the state $|\Psi\RR$.  The reduced state is
\begin{equation}
\rho_{12}(t)=\sum A_{\pm}(t)| \chi\RL\chi |
A_{\pm}(t)^{\dagger}. \;
\end{equation}
By the convexity of the entanglement of formation \cite{bennett},
$\mathcal{E}$$(\sigma)\geq \sum_{i} p_{i} $$\mathcal{E}$$(\sigma_{i})$ where the
state $\sigma_{i}$ is related to $\sigma$ by local operations and
$\sum p_{i}=1$.  Therefore, $\mathcal{E}$$(\rho_{12}(t))\leq $$\mathcal{E}$$(\rho_{12}(0))$.

\subsubsection{Bipartite entanglement between 2 and 3}

Consider the following initial state which is seperable over qubits 1 and 2
\begin{equation}
\begin{array}{lll}
| \Psi\RR&=&| \phi\RR_{1}| \chi\RR_{23}\\
&=&| \phi\RR_{1}(a|0\RR_{2}|0\RR_{3}+b|0\RR_{2}|0\RR_{3}). \;
\end{array}
\end{equation}
As before, we have expanded the entangled state between qubits 2 and 3
in a Schmidt basis with real parameters $a,b$.  This state is equivalent (up to
local rotations) to a linear combination of eigenstates of
$\sigma^{1}_{\vec{\alpha}}\otimes\sigma^{2}_{\vec{\beta}}
\otimes\sigma^{3}_{j}$ as
\begin{equation}
| \Psi\RR=a R_{1} \otimes R_{2} \otimes R_{3}|\displaystyle{++}+\RR
+b R_{1} \otimes R_{2} \otimes R_{3}|\displaystyle{+-}-\RR. \;
\end{equation}

Here a Krauss operator expansion is not particular helpful because the
two qubit system 1,2 is initially entangled with qubit 3.
However, the evolved three body state is easily found using the expansion
for $R_{3}$,
\begin{equation}
\begin{array}{lll}
U(t)| \Psi\RR&=&V_{1_{+}}(t)\otimes V_{2_{+}}(t)(ac| \displaystyle{++}\RR+bd^{*}
|\displaystyle{+-}\RR)_{12}|+\RR_{3} \\
& &+V_{1_{-}}(t)\otimes V_{2_{-}}(t)(ad|\displaystyle{++}\RR-bc^{*}|\displaystyle{+-}\RR)_{12}|-\RR_{3} \\
&=&\sum p_{\pm} V_{1_{\pm}}(t)\otimes V_{2_{\pm}}(t)|+\RR_{1}|u_{\pm}\RR_{2}|\raisebox{-.15ex}{\mbox{$\pm$}}\RR_{3}, \;
\end{array}
\end{equation}
where
\begin{equation}
 p_{+}=\sqrt{|ac|^{2}+|bd^{*}|^{2}},p_{-}=\sqrt{|ad|^{2}+|bc^{*}|^{2}}, \;
\end{equation}
and
\begin{equation}
\begin{array}{lll}
|u_{+}\RR&=&1/p_{+} (ac|+\RR+bd^{*}|-\RR), \\
|u_{-}\RR&=&1/p_{-} (ad|+\RR-bc^{*}|-\RR), \;
\end{array}
\end{equation}
are the normalized states for qubit 2.
The reduced state is then,
\begin{equation}
\rho_{12}(t)=\sum p_{\pm}^{2} V_{1_{\pm}}(t) \otimes V_{2_{\pm}}(t)
| \displaystyle{+u_{\pm}\RL+u_{\pm}}|
V_{1_{\pm}}(t)^{\dagger} \otimes V_{2_{\pm}}(t)^{\dagger}; \;
\end{equation}
a convex combination of seperable states which is therefore seperable:
$\mathcal{E}$$(\rho_{12}(t))=$$\mathcal{E}$$(\rho_{12}(0))\\=0$.
This argument is completely symmetric under interchange of 1 and 2 for the
case of initial bipartite entanglement between qubits 1 and 3.

\subsubsection{Three party entangled states}
The third set of initial pure states to consider is states that cannot be seperated over
any set less than order three.  There are several classifications of three party entangled states,
but we focus on a division imparted by a physical quantity
introduced by Coffman \textit{et.al.} \cite{coffman} known as the residual tangle $\tau_{123}$.
This quantity can be thought of as the amount of entanglement between 1 and the joint system
2,3 that cannot be accounted for by the entanglements of 1 with 2 and 3 seperately.  For qubits
such a quantity is naturally symmetric with respect to interchange of particles and is shown
to be equal to
\begin{equation}
\tau_{123}=2(\lambda^{12}_{1} \lambda^{12}_{2}+\lambda^{13}_{1} \lambda^{13}_{2}), \;
\end{equation}
where $\lambda^{1j}_{1}$ and $\lambda^{1j}_{2}$ are the square roots of the two eigenvalues of
$\rho_{1j} \tilde{\rho}_{1j}$.
A notable result of D\"{u}r \textit{et.al.} \cite{dur} shows that states with zero residual tangle
cannot be converted to GHZ states (states with $\tau_{123}=1$) under stochastic local operations
and classical communication (SLOCC).  They shown that all members of the set of zero residual tangle
(ZRT) states can be written,
\begin{equation}
| \Psi^{ZRT}\RR=a|000\RR+b|001\RR+c|010\RR+d|100\RR. \;
\end{equation}

Brun and Cohen \cite{brun} have constructed a distillation protocol using local POVM's that evolves all ZRT states to
a boundary of the set defined triple states
\begin{equation}
| \Psi^{tr}\RR=f|001\RR+g|010\RR+h|100\RR. \;
\end{equation}
This is one class of initial states we study under evolution by Eq. \ref{genstate}.  The second class we consider
is the set of generalized GHZ states,
\begin{equation}
| \Psi^{GHZ}\RR=a|000\RR+b|111\RR, \;
\end{equation}
which have residual tangle $\tau_{123}=4ab$ and equal the standard GHZ state when $a=b=1/\sqrt{2}$.  While these
are not the most general states with nonzero residual tangle, it is the class most easily distilled to
a GHZ state \cite{brun}.  The states $| \Psi^{tr}\RR$ and $| \Psi^{GHZ}\RR$ are inconvertible under
SLOCC.
\\
\\
a.  \textit{Generalized GHZ states}
\\
\\
Let
\begin{equation}
\begin{array}{lll}
| \Psi^{GHZ}\RR&=&a|000\RR+b|111\RR \\
&=&aR_{1}R_{2}R_{3}(a|\displaystyle{++}+\RR+b|\displaystyle{--}-\RR). \;
\end{array}
\end{equation}
Tracing over any one of the qubits leaves a mixed state for the other two with
no entanglement.  Under the action of $R_{3}$ (from Eq. 16),
\begin{equation}
| \Psi^{GHZ}\RR=R_{1}R_{2} ac|\displaystyle{++}+\RR+ad|\displaystyle{++}-\RR+bd^{*}
|\displaystyle{--}+\RR-bc^{*}|\displaystyle{--}-\RR. \;
\end{equation}
The evolved three body state is
\begin{equation}
\begin{array}{lll}
U(t)| \Psi^{GHZ}\RR&=&V_{1_{+}}(t) \otimes V_{2{+}}(t) (ac|\displaystyle{++}\RR+bd^{*}
|\displaystyle{--}\RR)_{12}|+\RR_{3}\\
& &+V_{1_{-}}(t) \otimes V_{2{-}}(t) (ad|\displaystyle{++}\RR-bc^{*}|\displaystyle{--}\RR)_{12}|+\RR_{3}\\
&=&\sum p_{\pm} V_{1_{\pm}}(t) \otimes V_{2{\pm}}(t)| \chi_{\pm}\RR_{12}|\pm\RR_{3}, \;
\end{array}
\end{equation}
where $p_{\pm}$ are the same as in Eq. 23 and
\begin{equation}
\begin{array}{lll}
| \chi_{+}\RR&=&1/p_{+} (ac|\displaystyle{++}\RR+bd^{*}|\displaystyle{--}\RR), \\
| \chi_{-}\RR&=&1/p_{-} (ad|\displaystyle{++}\RR-bc^{*}|\displaystyle{--}\RR),\;
\end{array}
\end{equation}
are the entangled states for the joint system 1,2.  The reduced state is then,
\begin{equation}
\rho_{12}(t)=\sum p_{\pm} V_{1_{\pm}}(t) \otimes V_{2{\pm}}(t)| \chi_{\pm}\RL\chi_{\pm}|
V_{1_{\pm}}(t)^{\dagger} \otimes V_{2{\pm}}(t)^{\dagger}.\;
\end{equation}

The evolved state is a statistical mixture of entangled states, which depending on the
weights of the mixture may yield greater than zero entanglement for the joint system 1,2.
Therefore $\mathcal{E}$$(\rho_{12}(t))\geq$$\mathcal{E}$$(\rho_{12}(0))$.
\\
\\
b.  \textit{Triple states}
\\
\\
Let
\begin{equation}
\begin{array}{lll}
| \Psi^{tr}\RR&=&a|001\RR+b|010\RR+k|100\RR \\
&=&R_{1}R_{2}(a|\displaystyle{++}\RR_{12}R_{3}|-\RR_{3}+b|\displaystyle{+-}\RR_{12}R_{3}|+\RR_{3}
+k|\displaystyle{++}\RR_{12}R_{3}|+\RR_{3}. \;
\end{array}
\end{equation}
The initial tangle between qubits 1 and 2 is $\tau(\rho_{12}(0))=4|bk|^{2}$.
After evolution, the state is,
\begin{equation}
\begin{array}{lll}
U(t)| \Psi^{tr}\RR&=&V_{1_{+}}(t) \otimes V_{2_{+}}(t)(ad^{*}|\displaystyle{++}\RR
+bc|\displaystyle{+-}\RR+kc|\displaystyle{-+}\RR)_{12}|+\RR_{3}\\
& &+V_{1_{-}}(t) \otimes V_{2_{-}}(t)(-ac^{*}|\displaystyle{++}\RR
+bd|\displaystyle{+-}\RR+kd|\displaystyle{-+}\RR)_{12}|+\RR_{3}. \;
\end{array}
\end{equation}
where we have used the rotation $R_{3}$ defined in \ref{rot}.  The reduced state is
\begin{equation}
\rho_{12}(t)=\sum m_{\pm}^{2} V_{1_{\pm}}(t) \otimes V_{2_{\pm}}(t)| \phi_{\pm}\RL\phi_{\pm}|
V_{1_{\pm}}(t)^{\dagger} \otimes V_{2_{\pm}}(t)^{\dagger}, \;
\end{equation}
where
\begin{equation}
\begin{array}{lll}
m_{+}&=&\sqrt{|a|^{2}+|c|^{2}-2|a|^{2}|c|^{2}}, \\
m_{-}&=&\sqrt{|a|^{2}+|d|^{2}-2|a|^{2}|d|^{2}}, \;
\end{array}
\end{equation}
and the normalized states for qubits 1 and 2 are
\begin{equation}
\begin{array}{lll}
| \phi_{+}\RR&=&1/m_{+} (ad^{*}|\displaystyle{++}\RR+cb|\displaystyle{+-}\RR+ck|\displaystyle{-+}\RR), \\
| \phi_{-}\RR&=&1/m_{-} (-ac^{*}|\displaystyle{++}\RR+db|\displaystyle{+-}\RR+dk|\displaystyle{-+}\RR). \;
\end{array}
\end{equation}

In order to relate the entanglement between qubits 1 and 2 before and after evolution,
we use the fact that the tangle is a convex function on the set of density matrices \cite{wootters}
such that
\begin{equation}
\begin{array}{lll}
\tau(\rho_{12}(t))&\leq&
\sum m_{\pm}^{2} \tau(|\phi_{\pm}\RR)\\
&=&m_{+}^{2} (4|cb|^{2}|ck|^2/m_{+}^{2})+m_{-}^{2} (4|db|^{2}|dk|^2/m_{+}^{2})\\
&=&\tau(\rho_{12}(0))(|c|^{4}+(1-|c|^{2})^{2})\\
&\leq&\tau(\rho_{12}(0)). \;
\end{array}
\end{equation}
Therefore $\mathcal{E}$$(\rho_{12}(t))\leq$$\mathcal{E}$$(\rho_{12}(0))$.

It is natural to ask what can be said of the residual
three tangle under this evolution.  Coffman \textit{et.al.} \cite{coffman} have constructed a
set of three quantities $(d_{1},d_{2},d_{3})$, invariant under permutation of three qubits,
that can be used to quantify the residual tangle as $\tau_{123}=4|d_{1}-2d_{2}+4d_{3}|$.
In terms of a state expanded
in the logical basis $|\xi\RR=\displaystyle{\sum_{ijk}} a_{ijk}|ijk\RR$, they are:
\begin{equation}
\begin{array}{lll}
d_{1}&=&a^{2}_{000}a^{2}_{111}+a^{2}_{001}a^{2}_{110}+a^{2}_{010}a^{2}_{101}
+a^{2}_{100}a^{2}_{011},\\
d_{2}&=&a_{000}a_{111}a_{011}a_{100}+a_{000}a_{111}a_{101}a_{010}+
a_{000}a_{111}a_{110}a_{001}\\
& &+a_{011}a_{100}a_{101}a_{010}+a_{011}a_{100}a_{110}a_{001}
+a_{101}a_{010}a_{110}a_{001},\\
d_{3}&=&a_{000}a_{110}a_{101}a_{011}+a_{111}a_{001}a_{010}a_{100}. \;
\end{array}
\end{equation}

Mapping $|+\RR\ra|0\RR, |-\RR\ra|1\RR$, it is clear from Eq. \ref{genstate} that the residual tangle
may either increase or decrease over time and will return to its initial value at times 
$|\vec{\alpha}|t=k\pi/2$ if $|\vec{\alpha}|/|\vec{\beta}|=k/l$ for some integer pair $k,l$.  The only invariant quantity above is $d_{3}$.
If one identifies the matrix elements
$a_{ijk}$ as vertices of a Boolean cube then $d_{3}$ is a sum of two configurations each lying on the
vertices of a tetrahedron \cite{coffman}.  These configurations are composed of four states with all even or odd Hamming weight.  For states with even or odd Hamming weight, $d_{1}$ and $d_{2}$ are zero and from the definition of $\tau_{123}$ the residual tangle is conserved under the evolution from commuting Hamiltonians with value 
$\tau_{123}=16|a_{000}a_{110}a_{101}a_{011}|$ or $\tau_{123}=16|a_{111}a_{001}a_{010}a_{100}|$.

\section{Noncommuting interactions}

In contrast to the case for commuting Hamiltonians, noncommuting pairwise interactions
can increase or decrease the entanglement between the noninteracting pair given any class of initial states.
To illustrate this we consider the Heisenberg interaction on the 1D chain of qubits oriented
(1,3,2) with equidistant spacing:
\begin{equation}
\begin{array}{lll}
H&=&\displaystyle{\sum_{<i,j>}}g_{ij}\sigma_{i}\cdot\sigma_{j}\\
&=&g\sigma_{1}\cdot\sigma_{3}+g\sigma_{2}\cdot\sigma_{3}\\
&=&H_{13}+H_{23}.\;
\end{array}
\end{equation}

There are three distinct eigenvalues of $H$ namely $(E_{0}=0,E_{1}=-4 g,E_{2}=2 g)$ corresponding
to eigenvectors antisymmetric under $1\lra2$, symmetric under $1\lra2$, and symmetric under all permutations
respectively.  Any state that is permutation symmetric, such as a tensor product of three identical states,
is stationary meaning the entanglement between
qubits 1 and 2 is invariant.  However, the state $|\Psi\RR=|00+\RR$, for instance, has overlap both with
eigenstates invariant under permutations of $1,2$ and $1,2,3$.  The reduced state of qubits 1 and 2 can be
easily be calculated by diagonalizing the Hamiltonian and
finding the Krauss operators $A_{0}={_{3}\LL}0|U(t)|+\RR_{3}$ and $A_{1}={_{3}\LL}1|U(t)|+\RR_{3}$ as,
\begin{equation}
\begin{array}{lll}
\rho_{12}(t)&=&A_{0}(t)|00\RL00|A_{0}(t)^{\dagger}+A_{1}(t)|00\RL00|A_{1}(t)^{\dagger}\\
&=&(2/9(1+\cos(6gt))+5/9)|00\RL00|+2/9(1-\cos(6gt))|\psi^{+}\RL\psi^{+}|\\
& &+\sqrt{2}/3 i\sin(3gt)e^{-i3gt}|00\RL\psi^{+}|-\sqrt{2}/3 i\sin(3gt)e^{i3gt}|\psi^{+}\RL00|.\;
\end{array}
\end{equation}
The tangle of this state is $\tau(\rho_{12}(t))=4/9 \sin^{3}(3gt)$.

For the other classes of states we do not present the calculations but merely state results.
For initial bipartite entanglement between 1 and 2, the tangle $\tau(\rho_{12}(t))$ can be either
greater or less than $\tau(\rho_{12}(0))$, while when the entanglement is initially shared between
1 and 3 (or 2 and 3), $\tau(\rho_{12}(t))\geq\tau(\rho_{12}(0))=0$.  The tangle is
zero for generalized GHZ states because these states are permutation symmetric.
The triple states can have $\tau(\rho_{12}(t))$ either be greater than or less than
$\tau(\rho_{12}(0))$.  It is notable that the residual tangle is invariant for the triple states
$(\tau_{123}=0)$ and the generalized GHZ states $(\tau_{123}=1)$.  This is a property of the
Heisenberg interaction having permutation symmetric eigenstates corresponding to these two classes.
Such symmetry persists even when the coupling constants $g_{ij}$ differ but is broken for any
anisotropy in the interaction.
For more general states the residual tangle can either increase or decrease under evolution.

\section{Conclusions}
We have investigated the preparation of entanglement between two qubits, 1 and 2,
that interact pairwise with a third qubit 3 but not with each other and in principle could be quite far apart from one another.
By explicit calculation we have found the change of entanglement between 1 and 2 for
various classes of initial states when the interaction Hamiltonians commute.  This is the case when the time order of the interactions is irrelevant.  Under such evolution, the entanglement of formation between 1 and 2 is nonincreasing for all initial pure states except the generalized GHZ states where it can increase from zero.  In contrast, noncommuting Hamiltonians can generally
increase or decease the entanglement between the noninteracting pair of qubits.  

One may be able to prepare entangled states from initial product states by performing a measurement on the third qubit.  This, in fact, is at the heart of the entangled chains proposal for quantum computing by Briegel and Raussendorf \cite{briegel} wherein a neighborhood of qubits perhaps confined to a lattice are allowed to interact by mutually commuting $\sigma^{i}_{z}\sigma^{j}_{z}$ interactions and arbitrary subsets can be prepared in entangled states by measurements on the remainder of the neighborhood.    In that proposal, because of the nature of the commuting interaction, measurement and local addressing of the qubits is necessary to entangle distant pairs.  Any entangling interaction can be mapping to any other two body entangling interaction by local unitary operations \cite{dodd}, however, if within a given system single qubit addressability is difficult this may not be possible.  A focus of future research will be to study what kind of entangled states could be synthesized given finite measurement resolution over sets of qubits interacting pairwise with each other.

\begin{acknowledgments}
I would like to thank Ivan Deutsch for many stimulating discussions.  The work was supported in part by a grant from the National Institutes of Standards and Technology (NIST) to the University of New Mexico (UNM) and by the National Security Agency (NSA) Advanced Research Development Activity (ARDA) under contract MOD7144.02. 
\end{acknowledgments}

\end{document}